\documentclass[aps, physrev, reprint, superscriptaddress]{revtex4-2}

\usepackage{amsmath, amssymb, amsthm, array, mathtools, xcolor, bm, bbm, hyperref, physics, tikz,algorithm,algpseudocode,qcircuit, textcomp}
\usepackage[nameinlink,capitalize]{cleveref}

\definecolor{darkblue}{RGB}{0,0,127} 
\definecolor{darkgreen}{RGB}{0,150,0}
\definecolor{tabred}{RGB}{214,39,40}
\definecolor{tabblue}{RGB}{31,119,180}
\hypersetup{breaklinks, colorlinks, linkcolor=darkblue, citecolor=darkgreen, filecolor=red, urlcolor=blue}
\hypersetup{pdftitle={}, pdfauthor={}}
\usetikzlibrary{patterns}
\usetikzlibrary{decorations.pathmorphing}
\usetikzlibrary{decorations.pathreplacing}
\include{figures/tensors}

\newcommand{\numberthis}{\addtocounter{equation}{1}\tag{\theequation}}

\newcommand{\expect}[1]{\mathbb{E}\left[#1\right]}
\newcommand{\isnorm}{\sim \mathcal{N}}
\newcommand{\T}{^\text{T}}
\newcommand{\argmin}[1]{\underset{#1}{\text{argmin\ \ }}}
\newcommand{\minimise}[1]{\underset{#1}{\text{minimise\ \ }}}

\newcommand{\acronym}{FBT}
\newcommand{\Eps}{\mathcal{E}}
\newcommand{\Cdd}{\mathbb{C}^{d\otimes d}}

\newcommand{\cA}{\mathcal{A}}

\newcommand{\cG}{\mathcal{G}}
\newcommand{\cU}{\mathcal{U}}
\newcommand{\xou}{U_{1,\uparrow}^{\pi/2}}
\newcommand{\xod}{U_{1,\downarrow}^{\pi/2}}
\newcommand{\xtu}{U_{2,\uparrow}^{\pi/2}}
\newcommand{\xtd}{U_{2,\downarrow}^{\pi/2}}

\newcommand{\zo}{Z_{1}^{\pi/2}}
\newcommand{\zt}{Z_{2}^{\pi/2}}

\newcommand{\favg}{F_\mathrm{avg}}

\newcommand{\tidlemid}{\raisebox{0.5ex}{\texttildelow}}

\definecolor{redg}{RGB}{224,102,102}
\definecolor{blueg}{RGB}{81,151,217}
\definecolor{greeng}{RGB}{106,168,80}
\definecolor{orangeg}{RGB}{250,160,50}
\definecolor{purpleg}{RGB}{140,125,195}

\DeclarePairedDelimiterX{\dsetting}[2]{\vert}{\rangle}{
	\mathopen{} #1 \mathclose{} \delimsize\vert \mathopen{} #2 \mathclose{} \delimsize\rangle\nhphantom{\rangle}{#2}
}

\DeclarePairedDelimiterX{\dangle}[1]{\langle}{\rangle}{
    \nhphantom{\langle}{#1} \delimsize\langle \mathopen{} #1 \mathclose{} \delimsize\rangle \nhphantom{\rangle}{#1}
}
\DeclarePairedDelimiterX{\dbra}[1]{\langle}{\vert}{
    \nhphantom{\langle}{#1} \delimsize\langle \mathopen{} #1 \mathclose{}
}
\DeclarePairedDelimiterX{\dket}[1]{\vert}{\rangle}{
    \mathopen{} #1 \mathclose{} \delimsize\rangle \nhphantom{\rangle}{#1}
}

\DeclarePairedDelimiterX{\dbraket}[2]{\langle}{\rangle}{
    \nhphantom{\langle}{#1} \delimsize\langle \mathopen{} #1 \mathclose{} \delimsize\vert \mathopen{} #2 \mathclose{} \delimsize\rangle \nhphantom{\rangle}{#2}
}

\newcommand{\nhphantom}[2]{\sbox0{$\left#1\vphantom{#2}\right.$}\!}

\DeclareMathOperator{\choi}{choi}
\DeclareMathOperator{\vecop}{vec}
\DeclareMathOperator{\multinomial}{Multinomial}
\begin{document}

\title{Fast Bayesian tomography of a two-qubit gate set in silicon}
\author{T.~J.~Evans}
\affiliation{Centre for Engineered Quantum Systems, School of Physics, The University of Sydney, Sydney, Australia}
\author{W.~Huang}
\author{J.~Yoneda}
\affiliation{School of Electrical Engineering and Telecommunications, The University of New South Wales, Sydney, NSW, Australia}
\author{R.~Harper}
\affiliation{Centre for Engineered Quantum Systems, School of Physics, The University of Sydney, Sydney, Australia}
\author{T.~Tanttu}
\author{K.~W.~Chan}
\author{F.~E.~Hudson}
\affiliation{School of Electrical Engineering and Telecommunications, The University of New South Wales, Sydney, NSW, Australia}
\author{K.~M.~Itoh}
\affiliation{School of Fundamental Science and Technology, Keio University, Yokohama, Japan}
\author{A.~Saraiva}
\author{C.~H.~Yang}
\author{A.~S.~Dzurak}
\affiliation{School of Electrical Engineering and Telecommunications, The University of New South Wales, Sydney, NSW, Australia}
\author{S.~D.~Bartlett}
\affiliation{Centre for Engineered Quantum Systems, School of Physics, The University of Sydney, Sydney, Australia}
\date{\today}

\begin{abstract}

Benchmarking and characterising quantum states and logic gates is essential in the development of devices for quantum computing.
We introduce a Bayesian approach to self-consistent process tomography, called fast Bayesian tomography (\acronym{}), and experimentally demonstrate its performance in characterising a two-qubit gate set on a silicon-based spin qubit device.  
\acronym{} is built on an adaptive self-consistent linearisation that is robust to model approximation errors.
Our method offers several advantages over other self-consistent tomographic methods.  Most notably, \acronym{} can leverage prior information from randomised benchmarking (or other characterisation measurements), and can be performed in real time, providing continuously updated estimates of full process matrices while data is acquired.  

\end{abstract}

\maketitle

\section{Introduction} \label{section:introduction}
In developing technology and devices for fault-tolerant quantum computing~\cite{gottesman1998ftqc}, the diagnosis and mitigation of errors in quantum gates is critical. 
A now-standard approach to assessing the performance of single- and two-qubit gates for quantum processors is to use randomised benchmarking (RB)~\cite{knill2008randomized,emerson2005scalable,magesan2011scalable,magesan2012characterizing} as well as variants built on this approach~\cite{dankert2009exact,magesan2012characterizing,granade2015accelerated,wallman2015estimating,dasilva2011practical,flammia2011direct,proctor2019direct}. 
The success of RB and related schemes comes from their efficiency for small systems as well as their robustness to state, preparation and measurement (SPAM) errors.  However, a key limitation is their inability, in general, to provide full diagnostic information about the types of errors.  This information is essential for these errors to be mitigated or eliminated.

When full characterisation of a gate or process is required, tomography remains the gold standard. 
Similar to RB, there are many methods to perform such characterisation, including tomography for unitary gates~\cite{kimmel2014robust,johnson2015demonstration} and for general channels~\cite{blume2017demonstration}.
An important step in the development of process tomography was the recognition that the operations used in the state preparation and measurement (SPAM) steps were also faulty, and required simultaneous characterization with the gates themselves. 
Process tomography that is capable of this simultaneous characterisation is said to be \emph{self-consistent}~\cite{merkel2013self}. Currently available implementations of self-consistent process tomography such as gate set tomography (GST)~\cite{blume2017demonstration} require the tailoring of specific sequences for specific gate sets. Typically, the number of sequences for two-qubit characterisation is very large, with commensurate computational requirements. 
The non-linear nature of tomographic reconstruction makes it difficult for non-expert users to understand the simplifying assumptions and technical insights required to make the numerical methods reliable in practice \cite{nielsen2020}, which may explain its limited adoption for characterizing two or more qubits.
Another limitation of all such systems is their inability to incorporate additional data or to update (without a full re-run of the non-linear, non-convex solvers) when more data become available.
Other tomography methods addressing some of these issues include Bayesian approaches~\cite{blumekohout2010optimal,schultz2017exponential,granade2016practical}, which provide the ability to encode prior information and so reduce the required data.
However, this efficiency is usually traded-off for computational cost due to the need for extensive sampling methods.

In this paper we present a new Bayesian method for self-consistent process tomography, called \emph{fast Bayesian tomography} (\acronym{}).
Our approach is able to extract tomographic information about a whole gate set from \emph{arbitrary} sequences of gates.
Our goal with \acronym{} was to create a tomography method that was operationally simple and efficient in both the experimental and computational cost.
The computational efficiency is achieved by defining a statistically optimal linearisation of the self-consistent objective function.
Moreover, the Bayesian model can be continuously updated as data becomes available, providing ongoing improvement to the linearisation. The data from standard randomized benchmarking experiments can be used to bootstrap the model. As we demonstrate, re-interpreting the RB data can already give good initial estimates of the process matrices of each of the gates, together with related uncertainties. Where required, such matrices and the uncertainties in their values can be updated with information from experiments as further data is extracted from the device.

To illustrate \acronym{} and its performance, we use it to characterise a 2-qubit gate set using the spins of electrons in a silicon-metal-oxide semiconductor (SiMOS) quantum dot device. Spin qubits in silicon are among the most promising semiconductor architectures for scalable quantum information processing due to their compatibility with modern semiconductor manufacturing process~\citep{Zwerver2021}, nanoscale footprint, and capability for hosting highly coherent qubits~\citep{Veldhorst2014,Muhonen2014}. 
High performance single-qubit gates are well characterised and have reached charge noise limits~\citep{yang2019silicon,Yoneda2018fidelity}. 
Moreover, two-qubit gates have recently demonstrated fidelities exceeding $99\%$~\citep{xue2021computing}.
Although some experiments have attempted to probe the type of noise affecting two-qubit gates~\citep{Xiao2019Fidelity,Boter2020Spatial}, the limiting factors remain unclear.  Identifying the key noise sources and failure modes for two-qubit gates lies at the heart of the development of semiconductor spin qubits, and it is here that a fast, flexible approach to self-consistent process tomography to characterise the noise is expected to be invaluable.

Specifically, we reconstruct a two-qubit native and non-standard gate set in a silicon spin qubit device, first demonstrated in Ref.~\citep{huang2019fidelity}, in a self-consistent manner.  We bootstrap data from an RB characterization of the gate set, followed by additional randomized gate sequences.  The results are high-accuracy estimates of the process matrices describing the gates, together with robust quantification of the uncertainties (error bars) on these estimates, highlighting the ability of the protocol to characterize non-standard gate sets as well as to leverage benchmarking data.  The processing of data takes less time than the data acquisition itself, demonstrating that \acronym{} can be performed in real time during an experiment.
Our Bell state tomography yields fidelities in the range  $94.6\%\mathrm{-}98.3\%$, comparable to those reported in Ref.~\cite{xue2021computing}, and concurrences from $88.3\%\mathrm{-}92.0\%$, the highest reported in silicon quantum dots. We find the individual two qubit conditional rotation (CROT) gate fidelity are between $96.1\% \mathrm{-} 97.4\%$, consistent with the gate fidelity reported in a previous experiment using the same device~\citep{huang2019fidelity}.  Finally, to illustrate the flexibility of \acronym{} for data that was not tailored specifically for tomography, we show that data from an RB characterisation of the device can be repurposed and can yield high-precision estimates of the gate set.

\Cref{section:results} presents an overview of the \acronym{} protocol and a summary of the experimental demonstration.
We then present a more detailed description and analysis of the \acronym{} protocol in~\cref{section:methods}.

\section{Results}
\label{section:results}

\subsection{FBT protocol}

In this section we outline the \acronym{} protocol and demonstrate its performance on a two-qubit gate set in a Si-MOS device.
\acronym{} provides a means of performing process tomography over a primitive gate set, in a self-consistent manner.
Moreover, it is efficient in its use of resources, both experimental and computational, and operationally simple, requiring no tailoring of complex experimental settings.

In an experiment involving a sequence of gates, the measurement outcomes are governed by Born's rule, and this provides us with the functional relationship between the gate parameters and our data.  Tomography is the procedure to invert this relationship to solve for the gate parameters given the measurement outcomes.  All tomographic methods suffer from the central problem that this function is a high-dimensional polynomial, of potentially high degree (for repeated gates).  Solving for the gates requires a nonlinear optimization that is typically intractable without simplifying assumptions, leading to large non-linear, non-convex optimisations that risk local-minima traps and can lead to long difficult computations~\cite{nielsen2020}.
For this reason, the initial proposal of self-consistent methods~\cite{merkel2013self} relied on linearisation to reduce the computational demand.
However, linearising around the identity imports a number of assumptions that can lead to poor performance. 

The Bayesian approach proposed in this paper gives us the ability to address this problem, and to do so in a way that is efficient in experimental and computational resources.
First, we show how to use prior information that we may already have about the gates, such as average gate fidelities from RB experiments, to provide an improved initial linearization.
Second, the Bayesian approach allows us to update our estimates of the gate set in real time (\emph{online}), as data becomes available.  Not only does this feature allow for data collection to be halted when reconstructions are of a desired accuracy, but also for additional data to be incorporated into the estimates without the need to re-run the entire computation.

Third, and most importantly, it provides us with an optimal linearisation.  Rather than linearising the model about the identity, \acronym{} instead uses the prior distribution as a starting point.  
This observation, albeit simple, has a significant impact on the utility and accuracy of the protocol.
The online nature of the protocol further supplements this feature as the mean is iteratively updated, improving the linearisation every time additional measurements are performed.
This also means that gate sets of arbitrary (e.g., lower) fidelity can be reconstructed, something many characterisation methods such as RB can struggle to do.
Moreover, the Bayesian model provides an accurate quantification of the approximation error incurred by the linearisation process.
As a result, we can take this error into account during the reconstruction. By knowing our estimate of the error we can prevent the model over-fitting to this error.

The Bayesian model uses a multivariate Gaussian prior over all of the selected channel parameters.
Each gate in the set is decomposed into the ideal implementation of a unitary gate $G$ followed by a gate-dependent noise channel $\Lambda_G$, such that $\tilde G = \Lambda_G G$. 
We are interested in reconstructing $\Lambda_G$ for each $G$, quantified by a mean $ \bar \Lambda_{G} $ and a covariance $ \Gamma_G $.  In principle, one would desire a prior that is constrained to be a physical quantum channel.  A Gaussian distribution does not necessarily guarantee a physical quantum channel and requires us to project onto one, but this limitation is far outweighed by the utility and robustness of the Gaussian model.

Full details of the \acronym{} protocol are given in~\Cref{section:methods}.
Here, we demonstrate \acronym{} on a gate set for a two-qubit device based on electron spins in silicon.

\subsection{Device and gate implementation}
\begin{figure}
	\centering
	\includegraphics[width=.9\columnwidth]{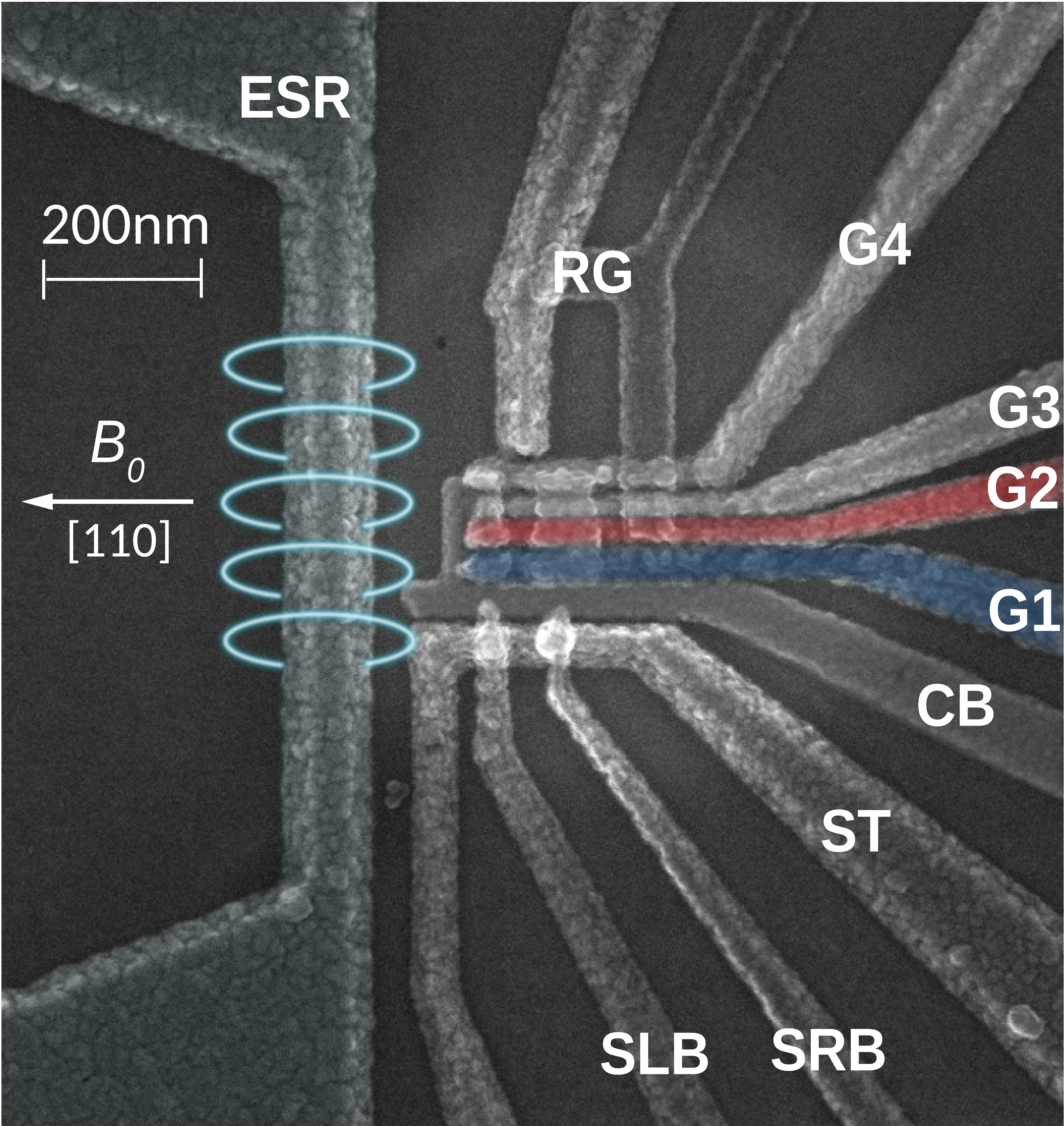}
	\caption{False color SEM image of the device. Two quantum dots are formed underneath gates G1 (blue) and G2 (red). The gates CB, G3 and G4 form confinement barriers that laterally define the quantum dots. RG is the reservoir gate that supplies electrons to the quantum dots. The gate electrodes ST, SLB and SRB define a single-electron transistor (SET), designed to sense charge movement in the quantum dot region. An alternating current running through the ESR line generates an oscillating magnetic field $B_1$ (light blue) to manipulate the electron spins. The direction of the external magnetic field $B_0$ is indicated by the white arrow.}
	\label{fig:ExerpiemntalSetup}
\end{figure}

\Cref{fig:ExerpiemntalSetup} shows a scanning electron microscope (SEM) image of a silicon-metal-oxide-semiconductor (Si-MOS) double quantum dot device, identical to the device used in this experiment. Qubits are defined in single electron quantum dots formed underneath gates G1 and G2. An external magnetic field $B_{0}=1.42$~T creates a Zeeman splitting of $E_\text{Z}=g \mu_\mathrm{B} B_{0}\approx 0.16$~meV, corresponding to an average electron spin resonance (ESR) frequency $f=E_\text{Z}/h=39.33$~GHz, where $g$ is the electron $g$-factor, $\mu_\mathrm{B}$ is the Bohr magneton and $h$ is Planck's constant. The device is operated in a dilution refrigerator at an electron temperature of $T_\text{e}\approx 180$~mK, allowing us to read the electron spin state sequentially via spin-dependent tunneling~\cite{Elzerman2004} and selectively load a $\ket{\downarrow}$ electron for initialisation. Single- and two-qubit control are enabled by a Zeeman energy difference of $\delta E_\text{Z}/h=13.42$~MHz and the exchange coupling $J/h=3.77$~MHz.

Two-qubit gates are commonly implemented in spin systems such as the $\sqrt{\text{SWAP}}$~\citep{Sigillito2019}, the controlled phase~\citep{Veldhorst2015,watson2018aprogrammable,Xiao2019Fidelity} or the controlled rotation (CROT)~\citep{huang2019fidelity,zajac439resonantly}. In our experiment, the 6 primitive gates (\cref{eqn:native-gate-set}) for universal qubit control consist of variations of CROT gates and virtual-Z gates. This choice of primitive gates allows us to control the qubits with constant exchange coupling, alleviating the requirement for high-bandwidth gate electrode voltage pulse and accurate synchronisation between signal sources.  High-fidelity one- and two-qubit gates were previously benchmarked on this device and reported in Ref.~\citep{huang2019fidelity}.

Our process tomographic characterisation will be at the level of the device's native primitive ($\pi/2$-pulse) gate set.
The gate set we consider is
\begin{align*}
\centering
&\Qcircuit @C=0.5em @R=0.6em {
	\lstick{} & \gate{X_{\frac{\pi}{2}}}  & \qw & & 
	\lstick{} &  \gate{X_{\frac{\pi}{2}}}  & \qw & & 
	\lstick{} & \gate{Z_{\frac{\pi}{2}}} & \qw & & 
	\\
	\lstick{} & \ctrl{-1}  & \qw &,& 
	\lstick{} & \ctrlo{-1}  & \qw &,& 
	\lstick{} & \qw & \qw &,&
	\\
	\\
	\lstick{} & \ctrl{1}  & \qw & & 
	\lstick{} & \ctrlo{1}  & \qw & & 
	\lstick{} & \qw & \qw& &
	\\
	\lstick{} & \gate{X_{\frac{\pi}{2}}}  & \qw &,& 
	\lstick{} &  \gate{X_\frac{\pi}{2}}  & \qw &,&
	\lstick{} & \gate{Z_\frac{\pi}{2}} & \qw& &
	{\inputgroupv{1}{2}{0.7em}{1em}{\cG = }}
	{\gategroup{4}{12}{5}{12}{1em}{\}}}
}
\end{align*}
which we will denote as
\begin{align}
\cG = &\{\xod,\xou,\zo,\xtd,\xtu,\zt\}\label{eqn:native-gate-set}
\end{align}
respectively.  See Ref.~\citep{huang2019fidelity} for further details about this gate set.

As mentioned, the single qubit $ Z $ gates are implemented by a virtual change of reference.
As a result, the fidelity of these gates can be expected to be significantly higher than the pulsed CROT-type gates.
Despite their virtual nature, we still consider the $Z$ gates as unknowns in the self-consistent protocol.

We also need to consider state preparations and measurements (SPAM), and the noise associated with them.
We assume that this can be modelled as a noise channel prior to the readout. 
The gauge degree of freedom between state preparation and measurement persists in this scenario, however based on early characterisation of this and similar devices it is reasonable to assume that the SPAM errors are dominated by the measurement.
This is limited by the signal strength as the qubits being read out are far away from the charge detector. Furthermore, we ensure the high initialization fidelity by choosing a readout time \tidlemid3.5~ms for both qubits which is sufficiently longer than the tunneling time of both ground and excited states.

\subsection{Randomised benchmarking}

We envisage our \acronym{} protocol being used immediately following an initial characterisation of the device using RB, and in particular \acronym{} is specifically designed to leverage the average gate fidelities from RB as prior information.  
We present the RB protocol used for the initial characterisation of the device, following Ref.~\cite{huang2019fidelity}.  In this protocol we generate gate sequences of varying lengths $L-1$, where all gates are randomly chosen from the two-qubit Clifford group. The final $L$-th Clifford gate at the end of each sequence is randomly chosen from those that would ideally return the final state to $\ket{\uparrow\uparrow}$.  The decay is fit to the function $P=A(1-\frac{4}{3}r_{\text{C}})^L+B$, where the fitting parameters $A$ and $B$ absorb the SPAM errors, and $r_{\text{C}}$ is the error per Clifford gate.  The resonance frequencies are re-calibrated after every 2 RB sequences.

From the decay, we extract an average Clifford gate fidelity of $F_{\mathrm{Clifford}}=1-r_{\text{C}}=90.5\pm 1\%$.
With the estimated average primitive fidelity $F_{\mathrm{\pi/2}}=98.2\pm0.2\%$ we can update our prior distribution to be consistent with this benchmark.
Since randomized benchmarking is designed to eliminate all coherent error and to average out incoherent error, we can only learn so much from the single parameter it returns. 
Applied to the whole gate set, this will mean the corresponding prior estimates for each noise channel will be pure depolarising channels.
The benefit of the RB protocol is that these estimates are unaffected by SPAM errors, and this benefit will assist in separating the SPAM errors from gate errors in the subsequent experiment.

\subsection{Self-consistent process tomography}
\label{subsection:results-fbt}

The initial characterisation using RB is used to construct nontrivial prior distributions for all of our gates based on the average gate fidelities from RB (see Sec.~\ref{subsection:methods-fbt-protocol}).  We then performed a series of tomography experiments. 
We measured 7140 settings consisting of sequences of primitive gates, with a frequency re-calibration after every sequence.
We randomly generated the set of tomographic settings of length $ L{=}0 $ (i.e., prepare and measure) to length $ L{=}14 $.  As we explain in the Methods,~\cref{section:methods}, \acronym{} is not prescriptive as to the sequences that need to be performed, but we note that we have chosen shorter-length sequences for tomography compared with the longer sequences used in RB.  
The circuits contained sequences of randomly-selected primitive gates in~\cref{eqn:native-gate-set}, and data acquisition took approximately 86 hours.
We note that this experiment goes well beyond what would traditionally be considered as an informationally-complete set of measurements.

\begin{figure*}[t!]
	\centering
	\begin{tikzpicture}
	\node at (0,0) {\includegraphics[width=\linewidth]{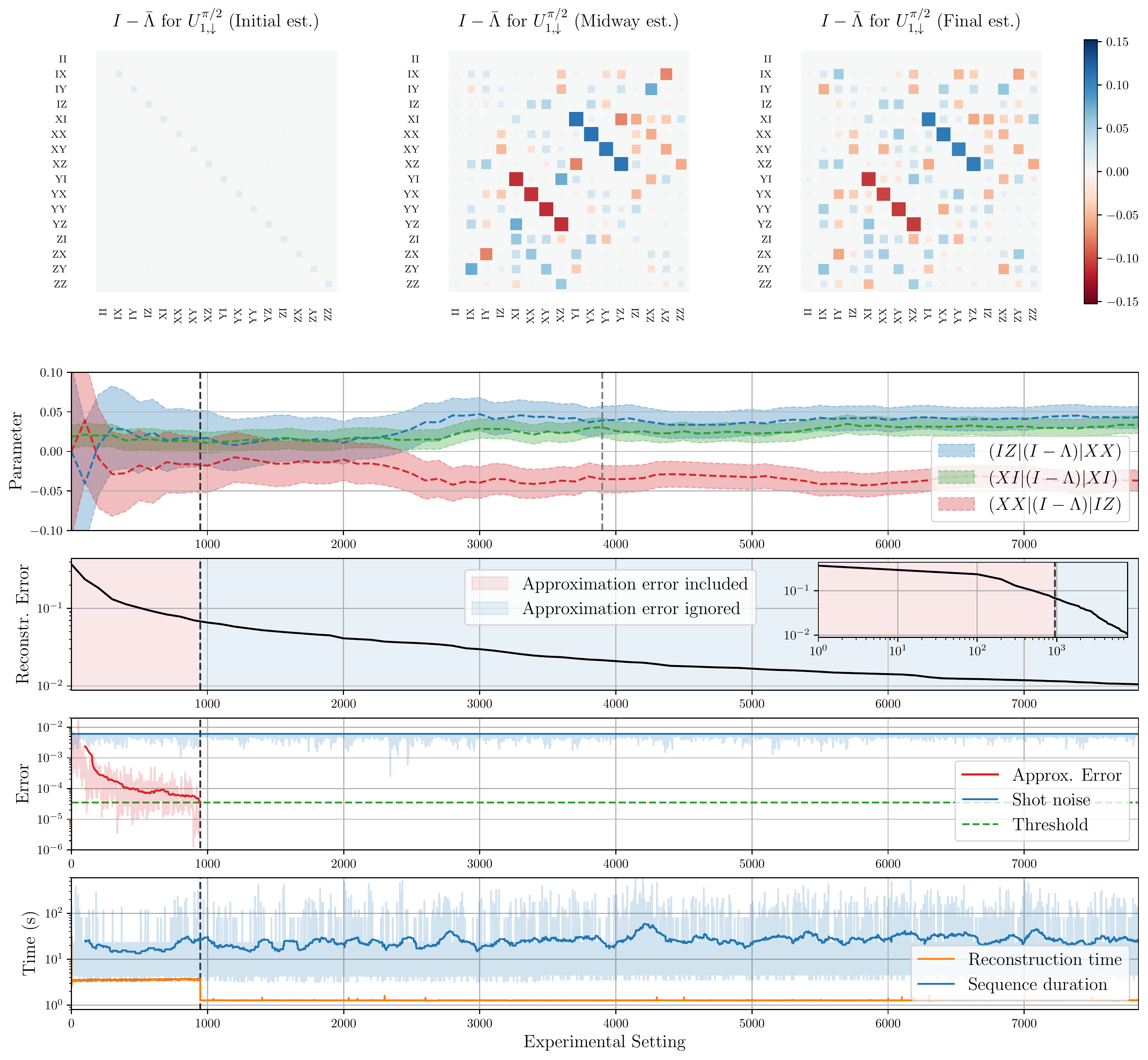}};
	
	\newcommand{\xinit}{-6.75}
	\def\sep{5.5}
	\def\xlist{\xinit,\xinit+\sep,\xinit+2*\sep}
	
	\newcommand{\width}{0.7}
    \newcommand{\ylower}{6.07}

    \foreach \x in \xlist{
        \draw [color=tabred, thick] (\x,\ylower) -- (\x,\ylower+\width/3) --(\x+\width*2/3,\ylower+\width) --(\x+\width,\ylower+\width) --(\x+\width,\ylower+\width*2/3) --(\x+\width/3,\ylower) --(\x,\ylower);
    }
    
	\draw [fill=white] (-8.1,7.75) circle (4pt) node[align=center] {1};
	\draw [fill=white] (-8.1+\sep,7.75) circle (4pt) node[align=center] {2};
	\draw [fill=white] (-8.1+2*\sep,7.75) circle (4pt) node[align=center] {3};
	
	\node at (-8.8,7.8) {\textbf{a)}};
	\node at (-8.8,2.6) {\textbf{b)}};
	\node at (-8.8,-0.3) {\textbf{c)}};
	\node at (-8.8,-2.9) {\textbf{d)}};
	\node at (-8.8,-5.4) {\textbf{e)}};
	
	\def\shadowtop{3.15}
	\def\shadowbottom{2.47}
	\def\shadowwidth{4}
    
    \def\base{-7.85}
    \fill [gray, opacity=0.2] (\base,\shadowbottom) --(-7.6,\shadowtop) --(-7.6+\shadowwidth,\shadowtop) -- cycle;

	\fill [gray, opacity=0.2] (0.4,\shadowbottom) --(-2.15,\shadowtop) --(-2.15 + \shadowwidth,\shadowtop) -- cycle;

    \fill [gray, opacity=0.2] (8.85,\shadowbottom) --(7.4,\shadowtop) --(7.4 - \shadowwidth,\shadowtop) -- cycle;

	\end{tikzpicture}
	\caption{High level schematic of the \acronym{} protocol and experiment, with the $\xod$ gate used as an illustrative example. 
	\textbf{a)} the estimated noise channel $\bar{\Lambda}$, presented as a Hinton diagram of the noise residual, $(I-\bar{\Lambda})$, for the $ \xod $ gate at three key stages during data acquisition:
	\protect\tikz \protect\node[circle,draw=black, fill=white, inner sep=0pt,minimum size=8pt] (0,0) {1};
	 the initial estimate based only on the fidelity;
    \protect\tikz \protect\node[circle,draw=black, fill=white, inner sep=0pt,minimum size=8pt] (0,0) {2};
	the estimate after half of the total number of sequences; and
	\protect\tikz \protect\node[circle,draw=black, fill=white, inner sep=0pt,minimum size=8pt] (0,0) {3};
the final estimate after all sequences have been measured.  The noise residual $(I-\bar\Lambda)$ reveals greater, clearer detail than the Pauli transfer matrix itself.
	\textbf{b)} the estimates for three parameters (indicated in \textbf{a)}), with mean values and uncertainties, as they are updated throughout the experiment. The three parameters were chosen only for clarity of presentation, having non-zero, non-overlapping mean values.
	\textbf{c)} the expected reconstruction error, as determined by the trace of the covariance for the gate, throughout the experiment. Inset:  as in the main plot, but using a logarithmic scale.  \textbf{d)} the magnitude of the two error processes -- the approximation error and shot noise -- throughout the experiment, characterised by the trace of their respective covariances. The approximation error is calculated as a moving average with window size of 100 (the darker line shows the moving average, the lighter lines the raw values).  A threshold for approximation error, at 2 orders of magnitude less than the shot noise, indicates when it can be neglected. For ease of reference this point is indicated by a vertical dotted line in all graphs b)--e).\textbf{e)} the speed of the \acronym{} protocol compared to the time to perform each sequence in the experiment.The light blue lines show the raw time of execution of each sequence (which varies with the length of any particular sequence), the darker line the moving average.}
	\label{fig:high-level}    
\end{figure*}

Results are presented in~\cref{fig:high-level}.  Between each setting, the measurement outcomes are used to update the posterior distribution of the gate sets.
This update then improves the linear model for the subsequent data.
Because we are using a linear approximation to the true nonlinear model, we end up with two sources of error: one from standard shot noise and the other from the linearisation.  The latter we refer to as the approximation error, and if this error is ignored (as in Refs.~\cite{merkel2013self, gu2020randomized}) it can lead to over-fitting, especially if the fitting function is linearised about the identity noise channel.
The \acronym{} protocol is robust to both of these error sources.
Once our estimates are accurate enough, the approximation error can become insignificant relative to the shot noise, and so if certain conditions are met we can omit the approximation error from the update which speeds up the inference significantly.  The point at which we omit the approximation error is illustrated in~\cref{fig:high-level}.  (We define these conditions more formally in \cref{section:methods}.)
This means the posterior distribution has contracted around the `true' gate set such that our linearisation performs as well as the exact nonlinear model, given the amount of shot noise in our measurements.  

For this experiment, the data were analysed after all measurements were taken.  Nonetheless, we can simulate the real time use of \acronym{}, and note that it is capable of updating estimates faster than measurements can be taken, despite not optimising the code for speed. \cref{fig:high-level}~\textbf{e)} shows that, on average, each sequence takes \tidlemid26~s. While the approximation error is being sampled, \acronym{} takes \tidlemid4~s to update for each sequence, and this reduces to \tidlemid1~s once the approximation error becomes negligible.
This affirms the potential for \acronym{} to be used in real-time while a device is being operated.  

\Cref{fig:high-level} shows a high level detail of the experiment and the corresponding reconstruction error throughout the experiment. 
The benefit of having a full Bayesian model for the gate set is that we always have a full covariance describing the uncertainty in the fits, including correlations between parameters as well as between separate gates.
The reconstruction error shown in~\cref{fig:high-level} is given by the mean square error 
of the estimate.
The log scale inset plot shows the rate at which the fits improve throughout the experiment.
As we expect, the mean square error approaches a central limit rate of \tidlemid$N_{\mathrm{settings}}^{-1}$, where $N_{\mathrm{settings}}$ is the number of experimental settings measured.
The full set of tomographic results for the gate set can be found in~\cref{appendix:gate-set-results}.

From the gate set estimates we can extract metrics and benchmarks of interest.
Gate dependent fidelities $f(G)$, unitarities $u(G)$, along with error bars, are all available to be extracted from the posterior distribution.
In place of the unitarity we present the gate dependent \emph{incoherence}~\citep{yang2019silicon}, $\omega(G)$, defined as
\begin{equation}
    \omega(G) := \tfrac{d-1}{d} \bigl(1 - \sqrt{u(G)}\bigr).\label{eqn:incoherence-def}
\end{equation}
The incoherence is bounded above by the infidelity, and gives a measure of how much of the gate infidelity can be attributed to incoherent errors and stochastic noise sources.
\Cref{figure:device-benchmarks}\textbf{a)} presents a summary of the gate-dependent primitive infidelities and incoherences extracted by \acronym{}.

\subsection{Bell-state tomography}

\begin{figure*}[t!]
	\centering
	\begin{tikzpicture}
	\node at (0,0) {\includegraphics[width=\linewidth]{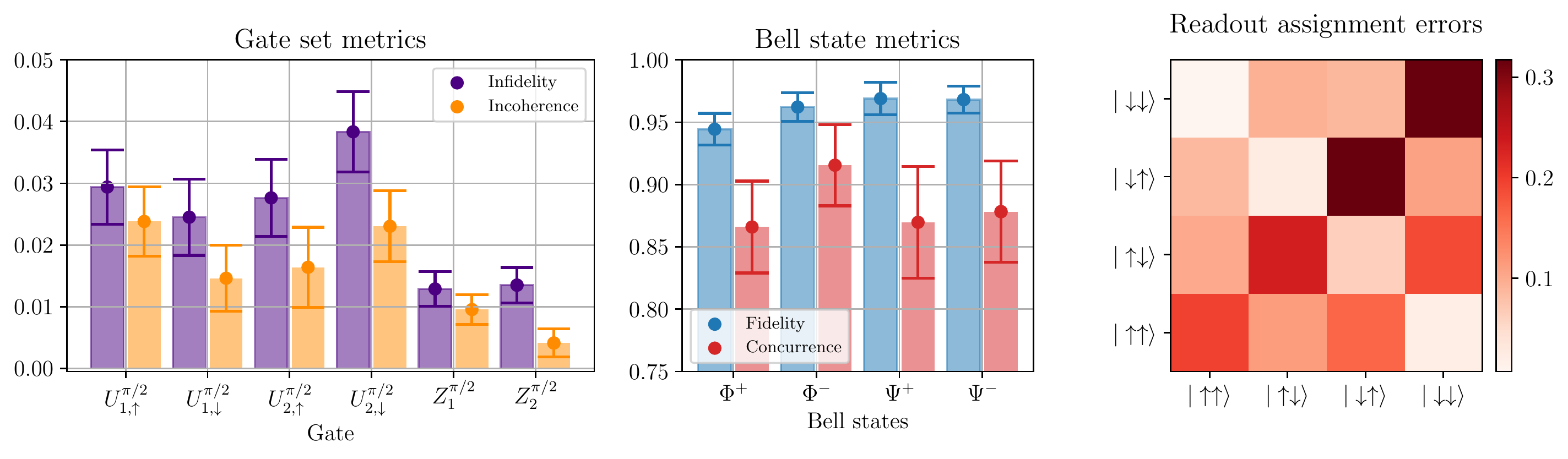}};
	\node at (-8.7,2.3) {\textbf{a)}};
	\node at (-1.5,2.3) {\textbf{b)}};
	\node at (3.7,2.3) {\textbf{c)}};
	\end{tikzpicture}
	\caption{Device benchmarks. \textbf{a)} Gate infidelity and incoherence for each individual primitive gate, extracted from the final posterior distribution. All error bars are 2 standard deviations which corresponds to a $95\%$-credible interval. \textbf{b)} Fidelity and concurrence for each of the 4 Bell-states $\Phi^{+},\Phi^{-},\Psi^{+},\Psi^{-}$. Having tomographic estimates for the primitive gates allows us to infer SPAM-free estimates of these benchmarks, compared to the conservative estimates in Ref.~\cite{huang2019fidelity} via state-tomography. \textbf{c)} Readout assignment matrix indicating assignment error rates for different preparations and measurements. The x-axis represents the prepared two-qubit state and the y-axis represents the corresponding measurement basis.}
	\label{figure:device-benchmarks}
\end{figure*}

Our \acronym{} protocol can also be adapted to perform state tomography self-consistently.  Although conceptually simpler than process tomography, state tomography generally requires the use of (noisy) gates to measure an informationally-complete set of observables.  Typically, state tomography is performed under the assumption that the uncharacterised, noisy gates are ideal, or that all measurements are described by a single `measurement fidelity.'  To perform state-tomography self-consistently, it is necessary to perform tomography on the gate set first.

Having performed a characterisation of the primitive gate set using \acronym{}, we can directly infer the states that can be generated in our device.
A common and important characterisation of a quantum processor is the quality of the entangled states that can be prepared.
\Cref{figure:device-benchmarks}\textbf{b)} presents the fidelities inferred from self-consistent state tomography for the four different Bell states, $\Phi^{+},\Phi^{-},\Psi^{+},\Psi^{-}$, generated from the same circuits as in Ref.~\cite{huang2019fidelity}.  These fidelities range from $94.6\%$ to $98.3\%$, with concurrences ranging from $88.3\%$ to $92.0\%$.
We note that these Bell-state fidelities and concurrences are larger than those presented in Ref.~\cite{huang2019fidelity}.  These improved metrics are not due to any performance improvement in the experiment; rather, they reflect improved statistical analysis as a result of using \acronym{}.

\subsection{Measurement tomography} \label{section:measurement-tomography}

A common approach~\cite{watson2018programmable, huang2019fidelity, yang2019silicon} to characterising readout assignment errors is to prepare and measure the various basis states by using gates to rotate between the desired basis state and the computational basis.
These assignment errors can be used to define a correction matrix which can applied to measurement data.
This approach, however, violates self-consistency as many of these measurements will also contain gate errors.

To complete the self-consistent characterisation of our device, we can add the measurement noise channel to the gate set, treating it as another random variable in the Bayesian model.
We model the SPAM as a single noise channel $\Lambda_E$ experienced by the state prior to readout, as in this device the readout errors dominate  errors in state initialisation. 
By including this measurement noise into the model, we are able to separate SPAM errors from gate errors.
This is aided by incorporating the RB inference into the prior. 
By providing the SPAM-free metric data into our model, we can more easily distinguish gate errors from SPAM errors in the subsequent experiments.
From the full SPAM noise channel $\Lambda_E$ we can infer the readout assignment error matrix for the spin states, shown in~\cref{figure:device-benchmarks}\textbf{c)}.

\subsection{Repurposing RB data}\label{subsection:results-repurposedRB}

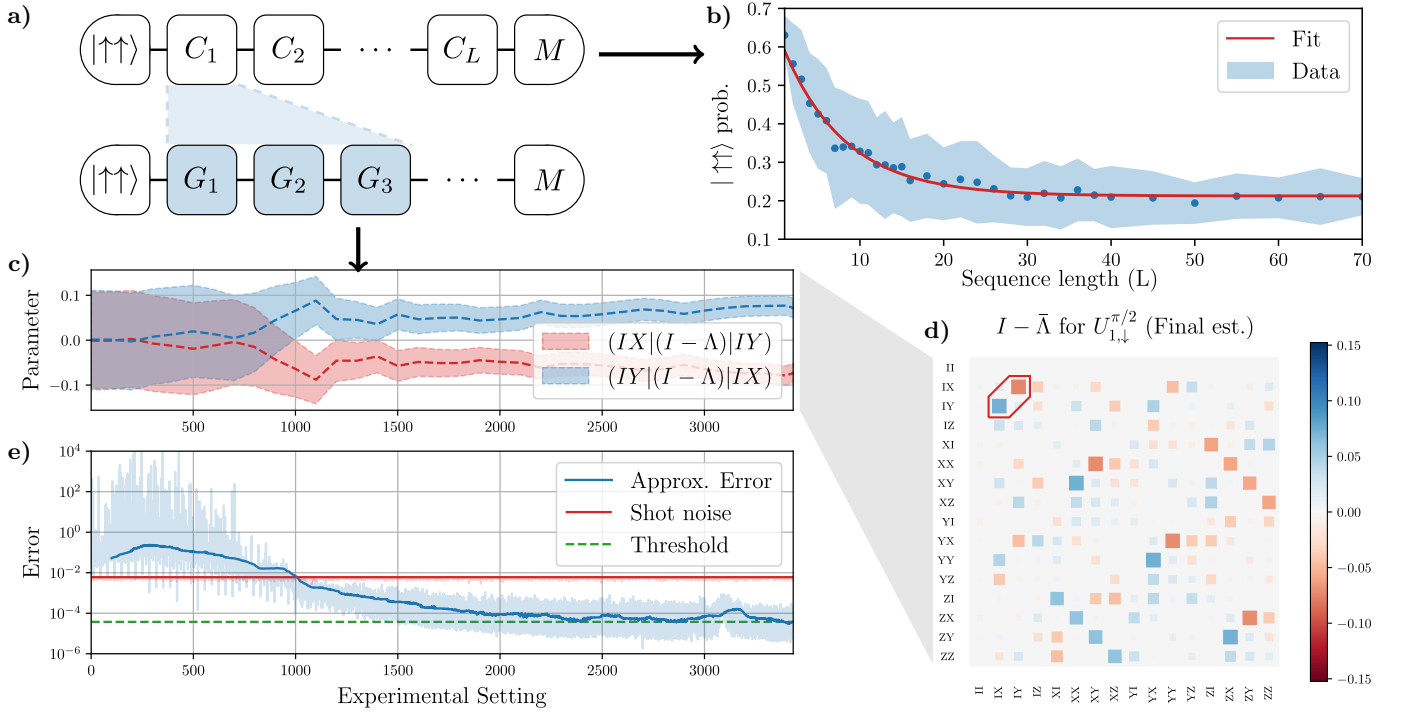
\begin{figure*}[t!]
	\centering
	\begin{tikzpicture}
	\node at (0,0) {\includegraphics[width=\linewidth]{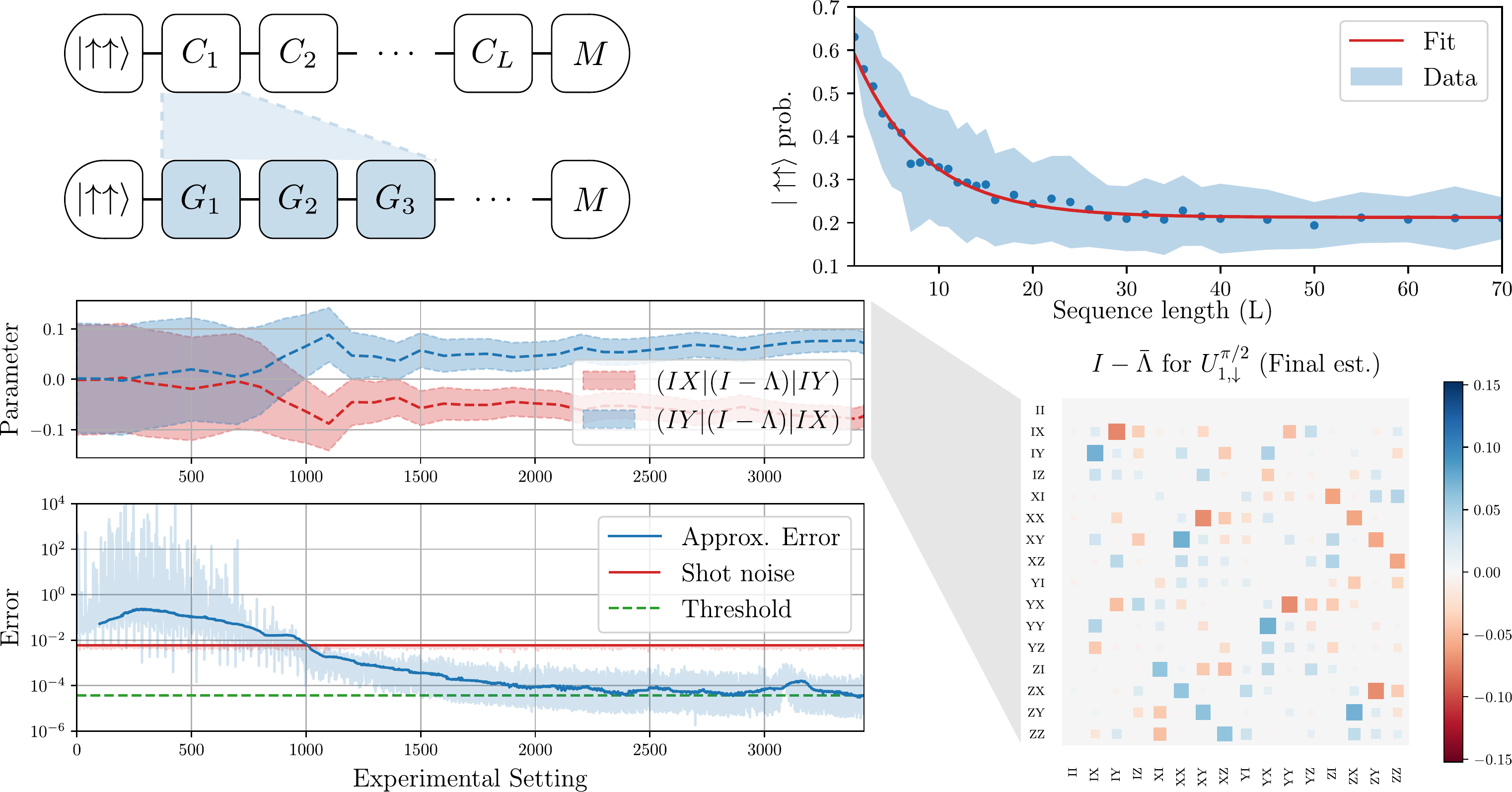}};
	\def\x{3.88}
	\def\width{0.55}
    \def\y{-0.83}
    \draw [color=tabred, thick] (\x,\y) -- (\x,\y+\width/2) --(\x+\width/2,\y+\width) --(\x+\width,\y+\width) --(\x+\width,\y+\width/2) --(\x+\width/2,\y) --(\x,\y);
	\draw [ultra thick, ->] (-1.3,4) --(0.1,4);
	\draw [ultra thick, ->] (-4.5,1.7) --(-4.5,1.1);
	\node at (-9,4.5) {\textbf{a)}};
	\node at (0.3,4.5) {\textbf{b)}};
	\node at (-9,1.2) {\textbf{c)}};
	\node at (-9,-1.3) {\textbf{e)}};
	\node at (3.2,0.3) {\textbf{d)}};
	\end{tikzpicture}
	\caption{Repurposing randomised benchmarking data for \acronym{}. \textbf{a)} the RB sequences can be repurposed as tomographic data for \acronym{} by decomposing Clifford gates $C_i$ into their primitive components $G_i$. \textbf{b)} the RB decay curve for the projected state probability as a function of the number of Clifford gates in each sequence. 
    \textbf{c)} the iterative estimates for two parameters (selected as examples and indicated in \textbf{d)}) throughout the RB experiment. \textbf{d)} the final estimate for $\xod$ gate, presented as a Hinton diagram of the noise residual $(I-\bar{\Lambda})$.  \textbf{e)} the magnitudes of the two error processes.  The approximation error is significantly larger in the repurposed RB sequences than the tomography data due to the use of much longer sequence lengths in RB. \label{fig:randomised-benchmarking}}
\end{figure*}

In the experiment reported above, we acquired new tomography data following the initial RB experiment in order to characterise the gate set.
One of the key benefits of \acronym{} is that it is able to extract tomographic information from any primitive gate sequence.
Since the RB sequences can be decomposed from Clifford circuits into sequences of primitive gates, we can bootstrap this data and perform \acronym{} over each primitive sequence as depicted in~\cref{fig:randomised-benchmarking}.
We can extract more information about the gates by asking more of our RB data than fitting only the usual decay curve, and in particular we can extract information about the coherent action of the noise present in the gates beyond the decoherent components that RB otherwise measures.

In this RB data, we have access to 3430 measurement settings ranging from primitive length of $1$ up to $1069$ primitive gates, all sampled for 125 shots.
\Cref{fig:randomised-benchmarking} shows the reconstruction of the $\xou$ gate throughout the experiment and we observe that we are able to reconstruct the process matrix to a moderate accuracy from the RB data alone.  We note that, due to the longer sequences used in RB, a longer period of data acquisition is needed before the approximation error becomes negligible.
Regardless, even when the approximation error is large, \acronym{} remains robust to it.
The fact that this was achieved with a standard RB dataset, from which a single average gate fidelity is usually all that is extracted, highlights the utility and flexibility of \acronym{}~\footnote{We note that, in principle, these data could have been combined with the tomography data in~\cref{subsection:results-fbt}.
In fact, using \acronym{} in this way would maximise what we can learn from RB data whilst supplementing it with shorter sequences to minimize the approximation error.  We have not combined our data sets here due to the fact that the RB and tomography data were taken at different times (separated by approximately one week), and the experiment may have suffered some drift.}.

\section{Methods}\label{section:methods}

Here, we provide details of the \acronym{} protocol and how it performs.

\subsection{Tomography and self-consistency} \label{subsection:problem-setting}

Consider a system consisting of $n$ qubits. 
Let $\mathcal{P} = \{P_i\}$ denote the set of all $n$-qubit normalised Pauli operators.
Suppose we have a set of $N_G$ gates $ \mathcal{G} = \{\tilde{G}_i \} $ that we can perform and that we are interested in characterising.

Each $ \tilde{G}_i\in \mathcal{G} $ is formally described by a completely positive trace preserving (CPTP) map.
We will work in the Pauli transfer matrix (PTM) representation, which as detailed in Ref.~\citep{kimmel2014robust}, has many useful features that we can take advantage of when parameterising our channels.
Standard quantum process tomography (QPT) attempts to reconstruct this matrix by performing an informationally complete set of measurements $\{M_i\}$ and input preparations $\{\rho_j\}$.
Given measurements of a gate $\tilde{G}$ of the form
\begin{align*}
    m_k &= \dangle{M_i | \tilde{G} | \rho_j}\\
    &= \dbraket{M_i\otimes\rho_j}{\tilde{G}}\\
    &=: A_{k} \dket{\tilde{G}} \numberthis\label{eqn:qpt-single-setting}
\end{align*}
where $\dket{\tilde{G}}\in\mathbb{C}^{d^2}$ denotes the $\vecop(\tilde{G})$ operation using the row-major order convention and $(i,j)\mapsto k$ is some labelling of the experimental settings.

If we stack our measurements such that
\begin{align*}
m&:= \begin{bmatrix}
m_1\\
\vdots\\
m_N
\end{bmatrix},\indent
A:=\begin{bmatrix}
A_1\\
\vdots\\
A_N
\end{bmatrix}
\end{align*}
then
\begin{align*}
    m = A\dket{\tilde{G}} \numberthis\label{eqn:qpt-all-settings}
\end{align*}
defines a linear inverse problem in the entries of $\dket*{G}$. 
The estimator for standard QPT is usually taken as the maximum likelihood,  given by
\begin{align*}
    \hat{G} &= \argmin{X\in\Cdd} \bigl\|m - A\dket*{X} \bigr\|_2^2\numberthis\label{eqn:qpt-mle}\\
    &= \left(A^\dagger A\right)^{-1} A^\dagger m. \numberthis\label{eqn:qpt-moore-penrose}
\end{align*}
where~\cref{eqn:qpt-moore-penrose} makes use of the Moore-Penrose pseudoinverse.

The shortcomings of this approach have been well explored: with high probability the estimator will not be a CPTP map; the estimated fidelity can disagree with estimates from RB; the assumption is made that measurement errors are i.i.d; etc.
However, the most concerning shortcoming is the lack of \emph{self-consistency}, introduced in Ref.~\cite{merkel2013self}.
This arises because the input states and measurements are generated by gates that have yet to be characterised, and which are themselves noisy.

Self-consistent process tomography addresses this problem by performing a simultaneous reconstruction of the entire gate set, including preparations and measurements.
Without loss of generality we can decompose each noisy gate $\tilde{G}_i = \Lambda_i G_i$ into the ideal gate $G_i$ followed by a noise channel $\Lambda_i$.
Then each experimental setting is described by a sequence of $N_k$ gates where
\begin{align*}
    m_k &= E \prod_{i\in S_k} \tilde{G}_{i} \dket{ \rho_0}\\
    &=E \prod_{i\in S_k} \Lambda_{i} G_{i} \dket{ \rho_0}\\
    &=:\mathcal{A}_k(\lambda)\numberthis\label{eqn:nonlinear-measurement-sequence}\,.
\end{align*}
Here, $S_k:\mathbb{Z}_{N_k}\to\mathbb{Z}_{N_G}$ is a tuple of indices corresponding to which gates were performed in the $ k^{\mathrm{th}} $ measurement setting and $\lambda:=\left[\dbra*{\Lambda_1},\dots,\dbra*{\Lambda_{N_G}}\right]^\dagger$ contains all of the unknown parameters of the noise channels.
The readout is a POVM $\{E_1,...,E_M\}$ (often, the set of $n$-qubit computational basis projective measurements) which can be combined in matrix form as
\begin{align*}
	E := \begin{bmatrix}
	\dbra{E_1}\\
	\vdots\\
	\dbra{E_M}
	\end{bmatrix}.
\end{align*}
In this self-consistent picture, there is no longer a single gate for which we collect tomographic data by varying the inputs and outputs.  Rather, each choice of experimental setting $m_k$ contains tomographic data of \emph{all} gates in that setting.

The tomographic reconstruction in this self-consistent scenario is much more challenging than just applying the simple formula of~\cref{eqn:qpt-moore-penrose}, with a non-linear inversion of a high-degree (with many settings) polynomial required for~\cref{eqn:nonlinear-measurement-sequence}.  This problem of non-linear inversion can be addressed by linearisation. The approach used in Ref.~\citep{merkel2013self} expresses each noise channel $\Lambda_i = I + \Eps_i$ and linearises~\cref{eqn:nonlinear-measurement-sequence} to obtain
\begin{align*}
	m_k &\approx E \prod_{i\in S_k} G_{i}\dket*{\rho_0}\\	
&+ \sum_{j=1}^{N_k} E \prod_{\substack{i\in S_k\\i<j}} G_{i} \Eps_{j} G_{j} \prod_{\substack{i\in S_k\\i>j}} G_{i} \dket{ \rho_0}\\
	&=: m_{\text{ideal}} + A_k x\numberthis\label{eqn:sc-model-merkel}
\end{align*}
where $m_\mathrm{ideal}$ is the ideal (noiseless) output and now $x=[\dbra*{\Eps_1},\cdots,\dbra*{\Eps_{N_G}}]^\dagger$.
This linear model yields a more efficient estimation of the gate set contained in $x$.
Computational speed is crucial so that insights from tomography can be used for error mitigation~\cite{white2019performance} before the noise in the system drifts.
However, by linearising about the identity $I$ for each channel $\Lambda_i$, this approach will only be accurate for gates with high fidelity (i.e., noise channels close to identity).
However, process tomography is predominantly employed to understand noise processes for the purpose of error mitigation, and demanding high fidelity gates is prohibitive in many experimental settings.
Ideally, process tomography should be suitable for use with gates of arbitrary fidelity for the purposes of early characterisation and tune-up.

To linearise the nonlinear function~\cref{eqn:nonlinear-measurement-sequence} about a better location requires some \emph{a priori} information about the noise in the gates.
In the next section we will describe how \acronym{} uses a Bayesian approach to construct a linear model that outperforms~\cref{eqn:sc-model-merkel}.
Moreover, our approach gives rise to an online model that can be updated as data becomes available, improving both the computational speed as well as the accuracy of the model.

\subsection{Fast Bayesian tomography (\acronym{})} \label{subsection:methods-fbt-protocol}

In this section, we formally introduce the \acronym{} protocol. 
Before we can define a linear model for \acronym{} we need to construct a Bayesian structure for the random variables involved.
Different prior models that reflect physical constraints have been previously proposed~\cite{blumekohout2010optimal,schultz2017exponential,granade2016practical,pogorelov2017experimentaladaptive,matteo2020operational}.
However, these approaches all lack scalability beyond single or two qubit processes as they rely on sampling methods such as sequential monte-carlo or importance sampling.
We instead propose the use of a Gaussian multivariate prior in order to pursue a conjugate posterior model that is computationally tractable and scalable.
Although this prior lacks the ability to constrain our parameter space to physical (CPTP) maps, we will show in~\cref{section:physical-priors} how this prior can still be used effectively to reflect physical constraints.

Firstly, let each noise channel $\dket{\Lambda_i}$ be distributed as a multivariate Gaussian $\dket{\Lambda_i}\isnorm(\dket{\bar{\Lambda}_i},\Gamma_i)$ with mean $\dket{\bar{\Lambda}_i}$ and covariance $\Gamma_i\in\mathbb{R}^{{d^2}\times{d^2}}$. 
In plain words, the mean $\bar{\Lambda}_i$ represents our best guess \emph{a priori} for the noise channel and similarly, the covariance $\Gamma_i$ encodes how much we trust the guess.
We will also abuse notation and interchangeably refer to both $\bar{\Lambda}_i$ and $\dket{\bar{\Lambda}_i}$ as the mean channel depending on the context.
We will also include SPAM noise channels $ \tilde{E}= E\Lambda_E $ and $ \tilde{\rho}_0=\Lambda_\rho \dket{\rho_0} $

One of the best parts of using Gaussian distributions is that we can efficiently sample from our prior.
We can likewise efficiently project a sample onto the convex set of CPTP maps.
We can combine these to inform a prior that will preclude channels that are very far from CPTP.
The goal here is to envelope the region of CPTP maps contained within the larger space of ${d^2 \times d^2} $ random matrices.
We will build on this in the next section, where we consider how we can incorporate prior information from benchmarking experiments, such as RB.

With prior knowledge about our gate set we have the ability to improve our model by linearising~\cref{eqn:nonlinear-measurement-sequence} about the mean, $\bar \lambda$.
This gives us the \acronym{} model
\begin{align*}
m_k\approx&E \bar\Lambda_E \prod_{i\in S_k} \bar{\Lambda}_{i} G_{i} \bar\Lambda_\rho \dket*{\rho_0 } \\
&+\sum_{j=1}^{N_k} E \bar\Lambda_E \prod_{\substack{i\in S_k\\i<j}} \bar{\Lambda}_{i} G_{i} \Eps_{j} G_{j} \prod_{\substack{i\in S_k\\i>j}} \bar{\Lambda}_{i} G_{i} \bar\Lambda_\rho \dket{ \rho_0}\\
=:& \bar{m}_k + \bar{A}_k x + e_k \numberthis\label{eqn:fbt-linear-model}
\end{align*} 
where $\bar{m}_k$ is the expected output of the setting, $\bar{A}_k$ is our linear model acting on the centralised variable $x$ and $e_k$ is the noise.
It is imperative in Bayesian inference to accurately quantify the noise in the measurement process as this informs the model of how strongly data should be weighted relative to the current prior information.
We have two sources of error in~\cref{eqn:fbt-linear-model}: 
the measurement noise due to finite averaging of experiment outcomes, and the model error incurred by the use of an approximate model.
We will treat both of these errors separately, $e_k:= \epsilon_k + \eta_k$ where $\epsilon_k$ and $\eta_k$ are the \emph{shot noise} and \emph{approximation error}~\cite{kaipio2007statistical}, respectively.

The approximation error is defined as
\begin{align}\label{eqn:approximation-error-def}
	\eta_k(x) := \cA_k (x+\bar\lambda) - \left(\bar{m}_k + \bar{A}_k x\right)
\end{align}
which we assume to be mutually independent of the statistical shot noise $\epsilon_k$.
The approximation error plays an important role when using a linearised model; omitting it can lead to over-fitting.
We can approximate both noise processes to be Gaussian as detailed in~\cref{appendix:bayesian-model} and it can be efficiently computed or directly sampled via our prior.
Having a Gaussian noise process and prior implies the posterior is also Gaussian, with mean and covariance that we can compute in closed form.

\acronym{} is an online protocol, meaning the posterior distribution over our gate set can be updated as measurements are received.
Throughout the protocol all we need to keep track of are our prior statistics $ (\bar{\lambda},\Gamma_{\lambda}) $.
When a new measurement $m_k$ is taken, we can compute our posterior distribution $x|m_k$ which immediately becomes the prior for the following measurement.
This means the approximate model is iteratively improved between every measurement setting.
\cref{appendix:bayesian-model} gives the explicit construction of this Bayesian model.

\subsection{Initialisation via benchmarking}\label{subsection:methods-initialisation-via-benchmarking}

One of the key benefits of having a Bayesian model is that we have the ability to include prior information, including benchmarks such as the average gate fidelity obtained from RB.
Although these benchmarks are typically limited in their ability to diagnose detailed noise processes, they do provide tight error bounds on gate set metrics.
The majority of randomised benchmarking experiments are also robust to SPAM.
This means that prior information extracted from benchmarking can inherit this SPAM-free property.
Also the more information we use prior to performing tomography, the better our approximate model will be, meaning less approximation error and a shorter time required before the approximation error can be ignored.
Finally, the random sequences performed in RB are suitable tomographic settings for \acronym{}, and can be reused as tomographic experiments, as we demonstrated in~\cref{subsection:results-repurposedRB}.

We briefly review the properties of data from standard RB~\cite{magesan2012characterizing}, which yields an average fidelity over our gate set obtained by fitting an exponential decay curve to the survival probability of Clifford circuits of increasing length.
Specifically, the average gate fidelity is
\begin{equation}
    \favg(\cG) = \frac{1}{N_G} \sum_{G\in\cG} f(G)\label{eqn:average-fidelity}
\end{equation}
where
\begin{equation}
    f(G) = \frac{\Tr(\Lambda_G) - 1}{d^2}\label{eqn:gate-fidelity}.
\end{equation}
The estimate obtained by this procedure will contain uncertainty, which we  take as our prior belief.
That is, we now consider the average gate fidelity to be a random variable, $f_\mathrm{avg}$ with some prior distribution.
The distribution we choose to encode this average fidelity prior may vary on how accurately (i.e., how many sequences and shots) the RB experiment was, and we only require that this distribution can be efficiently sampled.
Provided sufficient samples are taken, the distribution of $f_\mathrm{avg}$ should be well described by a univariate Gaussian distribution with mean $\bar f$ and variance $\sigma_f^2$.

With this prior information from RB, we can now construct priors for each gate in our gate set.  Given a sample of our gate set $\cG = \{G\}$ from the prior (which is not necessarily physical) we can take the projection onto the set of CPTP maps with a given average gate fidelity $f$ by the following SDP
\begin{align*}
    P_f(\cG) = \argmin{\{X_i\}} \sum_i \|\Lambda_i - X_i\|\\
    \text{subject to\indent} \favg(\{X_i\}) = f\\
    \choi(X_i) \succeq 0.\numberthis\label{eqn:sdp-gate-set-projection}
\end{align*}
\cref{alg:rb-update} details how we can create a prior distribution that envelopes the set of CPTP maps such that the average fidelity statistics are consistent with results obtained via RB.
\begin{algorithm}[H]
	\caption{\label{alg:rb-update} $\textsc{RB Update}(\bar{x}, \Gamma_x, \bar{f}, \sigma_f^2, N_\mathrm{samples})$\\ randomised Benchmarking prior update}
	\begin{algorithmic}[1]
		\Require $\bar{x}, \Gamma_x, \bar{f}, \sigma_f^2$\\
		\# $ x\isnorm\left(\bar{x},\Gamma_x\right) $ is the initial prior distribution\\
		\# $f_\mathrm{avg}\isnorm\left(\bar{f},\sigma_f^2\right) $ RB average fidelity prior distribution\\
		\For{$i$ in $1,...,N_\mathrm{samples}$}
		\State Sample an average fidelity $f$
		\State Sample a gate set $\mathcal{X}$
		\State Store the projected sample $P_f(\mathcal{X})$
		\EndFor
		\State take $(\bar{x},\Gamma_x)$ to be the sample mean and covariance of the projected samples
		\Ensure $(\bar{x},\Gamma_x)$
	\end{algorithmic}
\end{algorithm}
\cref{alg:rb-update} provides a mechanism to inject the average fidelity of a gate set into the \acronym{} protocol. We should note that other benchmarks, such as unitarity~\cite{wallman2015estimating}, can be used similarly, by adapting the projection in~\cref{eqn:sdp-gate-set-projection} accordingly.

\subsection{Including SPAM Errors} \label{section:mehtods-spam}

SPAM errors are central issue in characterising quantum devices.
However, there is a distinction that should be made between \emph{intrinsic SPAM errors}~\cite{blume2017demonstration}, i.e., errors in readout and state initialisation, and \textit{gate SPAM errors} that result from the imperfect gate used when, for example, preparing a specific state.
Self-consistency takes care of gate SPAM errors, since this is completely captured by the modelling the noise on the gates.
The intrinsic SPAM errors can be modelled as a noise channel on each of the state and measurement, i.e., $\tilde E = E \Lambda_E $ and $ \dket{\tilde\rho} = \Lambda_\rho \dket{\rho}$.  However, we note that there is a gauge degree of freedom between these two operators.
Theoretically, these two operators can be distinguished by using ancilla qubits~\citep{lin2019independent}.
In such a setting, this could then be used to inform separate prior distributions on each of the noise channels $\Lambda_\rho$ and $\Lambda_E$.
This is, however, beyond the scope of this paper (as well as most other self-consistent tomography methods).

Measurement errors are often characterised by estimating error rates for all computational basis states and measurements~\citep{yang2019silicon,huang2019fidelity,watson2018programmable}, which implicitly assumes that the measurement errors are classical.
Moreover, this approach requires the use of gates to initialise different input and output states, meaning the corresponding measurement fidelities will include gate noise, leading to unreliable estimates.
By including a measurement noise channel to the underlying model $\tilde E = E \Lambda_E $, this additional noise channel is added to the inference problem and kept separate from gate errors in our model.  An advantage of this approach is that it allows for the reconstruction of a quantum noise channel on our measurement, provided we can make assumptions about the state preparation as in~\cref{section:measurement-tomography}.

\section{Discussion}\label{section:discussion}

Fast Bayesian tomography is an efficient, flexible method for self-consistent tomography, with many features that make it appealing for the rapid early characterisation of quantum gates.  It is designed to be used immediately following RB to provide immediate diagnostic information about the gate performance and associated errors, allowing for rapid decisions on how gate fidelities can be improved.  Gates that have initially low fidelity are not a barrier as in previous methods.  The ability to incorporate prior information, such as the physics of state preparation and measurement, means that the gates are automatically constructed in a gauge familiar to those working with the device, without the need for any artificial gauge optimisation.  \acronym{} is very flexible in terms of the data that is uses; as we have shown, it can repurpose RB data, and it can be run on data from GST or other randomised sequences.  When new data are available, these can be directly incorporated into \acronym{} to provide further improved estimates.  We believe that \acronym{} fills an existing gap in the broad suite of characterisation tools, complementing more prescriptive benchmarking methods such as GST.

Our demonstration using a two qubit spin qubit device in silicon shows the utility of \acronym{}.  While this device was previously characterised using RB, we have shown how \acronym{} can bootstrap this RB data to provide the fidelity, unitarity, and full process matrices for all primitive 1- and 2-qubit gates.  The accuracy of our tomographic reconstructions is very high, with the approximation error due to our model becoming negligible.  The Bell state fidelities measured using \acronym{} are higher than the conservative estimates in Ref.~\citep{huang2019fidelity}, because of the self-consistent nature of \acronym{}.  Although we do not explore it further here, the detailed diagnostic information about the 2-qubit gates in this device may potentially be used to optimise the pulses and gate operating points such as to increase gate fidelities further.

\acknowledgments

We thank Steve Flammia, Thomas Smith, Owen Dillon, Thaddeus Ladd, and Matthew Otten for helpful discussions. This work was supported by the Australian Research Council via EQUS project number CE170100009, CQC2T project number CE170100012, and Laureate Fellowship project number FL190100167. RH acknowledges support from the Sydney Quantum Academy. We acknowledge support by the NSW Node of the Australian National Fabrication Facility. We also acknowledge support from the US Army Research Office grant number W911NF-17-1-0198. The views and conclusions contained in this document are those of the authors and should not be interpreted as representing the official policies, either expressed or implied, of the Army Research Office or the US Government.


%

\appendix
\filbreak
\setcounter{equation}{0}
\section{Experimental setup} \label{appendix:experimental-setup}

The quantum dots are formed in a $900$~nm thick isotopically enriched $^{28}$Si epi-layer (residual $^{29}$Si concentration of 800 ppm~\cite{itoh_watanabe_2014}) on a natural silicon substrate. They are laterally confined by aluminium gate electrodes fabricated using multi-layer gate stack technology. All measurements were performed in an Oxford Instruments wet dilution refrigerator with base temperature $T_\text{bath}\approx30$~mK and electron temperature $T_\text{e}\approx 180$~mK.
Battery-powered voltage sources (Stanford Research Systems, SIM928) and an arbitrary waveform generator (LeCroy ArbStudio, 1104 AWG) were used to generate DC voltages and voltage pulses, respectively, which were added through combiners with attenuation ratios 1:5 for DC voltages and 1:25 for voltage pulses. Low-pass filters were included for slow and fast lines (10 Hz to 80 MHz). An Agilent E8267D microwave vector signal generator was used to deliver ESR pulses to the on-chip microwave antenna after being attenuated at the 1.5~K stage (10~dB) and the 30 mK stage (3~dB). We used the internal arbitrary waveform generator of E8267D to perform single side-band modulation (by mixing the carrier with the in-phase and quadrature signals) and generate the four different ESR-drive frequencies.

\section{Bayesian model for \acronym{}} \label{appendix:bayesian-model}

\subsection{Construction of the likelihood} \label{subsection:methods-construction-likelihood}

We want to construct a Bayesian model that will give an updated estimate for our gate set given a new measurement $m$. 
In general, $m$ could be a set of measurements but as we are only interested in running the \acronym{} protocol online, it suffices to consider the case where $m$ is the measuement outcome from a single experimental setting.
This amounts to determining the conditional probability distribution of $x$ given $m$, known as the \emph{posterior} distribution which we will denote as $\pi(x|m)$.
By Bayes theorem we have
\begin{align}
    \pi(x|m) \propto \pi(m|x) \pi(x)\label{eqn:bayesian-posterior}
\end{align}
where $\pi(m|x)$ and $ \pi(x)$ are the likelihood and prior distributions respectively. 
We will begin by determining the likelihood for our approximate model.

Firstly, note that the linearisation that we have in~\cref{eqn:fbt-linear-model} is affine in our gates, and linear in $x$ containing the parameters of the centralised random variables $\dket{\Eps_i}$.
We begin by assuming that our random variables $m,x$ and $e$ are jointly normal with probability density $\pi(m,x,e)$.
Through repeated use of Bayes' theorem observe
\begin{align*}
    \pi\left(m,x,e\right)&= \pi\left(m|x,e\right)\pi\left(e|x\right)\pi\left(x\right)\\
    &= \pi\left(m,e|x\right)\pi\left(x\right)
    \intertext{hence}
    \pi\left(m,e|x\right) &= \pi\left(m|x,e\right)\pi\left(e|x\right). \numberthis \label{eqn:repeated-bayes}
\end{align*}

Using our model, the distribution of our data $m$ \emph{given} both the unknown gate parameters $x$ and the noise $e$ is completely determined.
This means
\begin{align}
    \pi(m|x,e) = \delta(-\bar{m} - \bar{A} x - e). \label{eqn:delta-distribution-data}
\end{align}
Hence, combining~\cref{eqn:repeated-bayes} and~\cref{eqn:delta-distribution-data} we can marginalise over the noise to get our likelihood
\begin{align*}
    \pi(m|x) &= \int\pi(m|x,e)\pi(e|x) d e \\
    &= \int \delta(-\bar{m} - \bar{A} x - e) \pi(e|x) d e\\
    &= \pi_{e|x} (-\bar{m} - \bar{A} x | x).
\end{align*}
The above distribution $\pi_{e|x} (-\bar{m} - \bar{A} x | x)$ is the conditional distribution of $e|x$. Usually noise in estimation problems can be assumed to be mutually independent of the unknown parameters, however the approximation error $\eta$ depends explicitly on our prior. 

Hence, we have the likelihood
\begin{align}
    m|x\isnorm(m - \bar A x - \bar e|x,\Gamma_{e|x}).
\end{align}
In the next section we will compute the statistics of this conditional noise process $e|x = \epsilon|x + \eta|x$.

\subsection{Noise processes}\label{subsection:methods-noise}

In our model we have two separate noise processes: the shot noise and the approximation error.
To compute the distribution of the noise consider the joint distribution
\begin{align*}
\pi (x,\eta,\epsilon) & \propto \\ 
\mathrm{exp} \Bigg( -\frac12 & \begin{bmatrix}
x-\bar x\\
\eta - \bar \eta\\
\epsilon - \bar \epsilon\\
\end{bmatrix}\T
\begin{bmatrix}
\Gamma_x & \Gamma_{x \eta} & \Gamma_{x\epsilon}\\
\Gamma_{\eta x} &\Gamma_{\eta} & \Gamma_{\eta \epsilon}\\
\Gamma_{\epsilon x} & \Gamma_{\epsilon \eta} & \Gamma_{\epsilon}\\
\end{bmatrix}^{-1}
\begin{bmatrix}
x-\bar x\\
\eta - \bar \eta\\
\epsilon- \bar \epsilon\\
\end{bmatrix}\Bigg)\\
\numberthis\label{eqn:noise-joint-dist}
\end{align*}
where each block in the joint covariance denotes the corresponding covariance matrix, i.e. $\Gamma_{\eta x} = \expect{(\eta - \bar\eta)(x-\bar x)\T}$. Recall that we are interested in determining the distribution of $ e|x = \epsilon|x + \eta|x$. 

Using standard identities for Gaussian conditional distributions~\cite{kaipio2007statistical} we can show 
\begin{subequations}
\begin{align}
    e|x \isnorm\left(\bar e|x,\Gamma_{e|x}\right)\label{eqn:e-cond-x-distribution}
\end{align}
where
\begin{align}
    \bar e|x = \bar \eta + \bar \epsilon + \left(\Gamma_{\eta x} + \Gamma_{\epsilon x}\right) \Gamma_{x}^{-1} x
\end{align}    
and
\begin{align}
    \Gamma_{e|x} = \Gamma_\eta + \Gamma_\epsilon - \left(\Gamma_{\eta x} + \Gamma_{\epsilon x}\right) \Gamma_{x}^{-1} \left(\Gamma_{x\eta} + \Gamma_{x\epsilon}\right).
\end{align}
\end{subequations}

Fortunately, all these distributions are easily and efficiently sampled due to being multivariate Gaussians. 
This can be achieved by sampling from the prior and directly computing the corresponding samples of $\eta$ and $\epsilon$ using the exact and approximate forward models $\cA(x)$ and $\bar A x$ which we detail further in the next section.
This gives us access to the full joint distribution $\pi(x,\eta,\epsilon)$ in~\cref{eqn:noise-joint-dist} from which we can compute all of the necessary statistics for the likelihood. 

We say that the shot noise \emph{dominates} the approximation error~\citep{kaipio2007statistical} if
\begin{align}
    \Tr(\Gamma_{\epsilon}) \gg 	\Tr(\Gamma_{{\eta}}) + \|\bar{\eta}\|_2^2 .\label{eqn:aproximation-error-dominates}
\end{align}
Initially, with little or no prior information, this is will not be the case, especially for long sequences which we see in~\cref{fig:randomised-benchmarking}~\textbf{e)}. 
If~\cref{eqn:aproximation-error-dominates} is not satisfied, it is imperative that the approximation errors are quantified and accounted for.
However, if the shot noise does dominate the approximation error, then the approximation is accurate enough for the approximation error to be neglected.
This means that our prior is `close enough' to the true gate set that the linear model, relative the statistical shot noise, is as good as the nonlinear model.

The measurement noise $\epsilon$ in our model comes from the nature of quantum measurement.
Every setting requires $N$ repetitions of the same experimental setting: initialise $\rho_0$, apply the sequence of gates from $S$, readout and collate the output of each run.
In this way our data is inherently multinomial, where the bias probabilities are determined by the Born probabilities of the POVM operators $p = \left[\Tr(E_1 \rho),\dots,\Tr(E_M \rho)\right]$.
Using central limit theorem arguments we can assume that the corresponding noise $ \epsilon $ is well approximated as a multivariate Gaussian random variable with mean zero, provided $N$ is large enough and the device isn't completely noiseless.

We make this assumption for the two qubit gate set results shown in~\cref{section:results}.
This simplifes the computation of the noise as only the approximation error needs to be computed and we instead have $e|x \approx \epsilon + \eta|x$.
The multinomial covariance is given by
\begin{align}
	\Gamma_{ij}=\begin{cases}
	N  p_i (1-p_i) &i=j\\
	-N p_i p_j &i\neq j .
	\end{cases}\label{eqn:multinomial-covariance}
\end{align}
To estimate the Born probabilities we can average the outputs and take $p\approx m$.
This means that the noise process $ \epsilon\isnorm(0,\Gamma_{\epsilon}) $ where
\begin{align}
\Gamma_{\epsilon} = \frac{\Gamma}{N}\label{eqn:shot-noise-covariance}
\end{align}
from~\cref{eqn:multinomial-covariance}. 

Also, once we have the posterior has contracted sufficiently such that the shot noise dominates the approximation error we can omit it from the noise model completely.
This also means that we no longer need to sample any noise statistics and $e|x = \epsilon$, significantly speeding up the protocol as shown in~\cref{fig:high-level}.

\subsection{Computing the noise statistics via sampling} \label{subsection:sampling-noise-statistics}

In this section we will detail how the noise statistics can directly estimated via sampling.
Recall from~\cref{eqn:e-cond-x-distribution} that we have
\begin{align*}
    \bar e|x = \bar \eta + \bar \epsilon + \left(\Gamma_{\eta x} + \Gamma_{\epsilon x}\right) \Gamma_{x}^{-1} x
\end{align*}    
and
\begin{align*}
    \Gamma_{e|x} = \Gamma_\eta + \Gamma_\epsilon - \left(\Gamma_{\eta x} + \Gamma_{\epsilon x}\right) \Gamma_{x}^{-1} \left(\Gamma_{x\eta} + \Gamma_{x\epsilon}\right).
\end{align*}
As mentioned in~\cref{subsection:methods-noise}, we can sample from the full joint distribution $\pi(x,\epsilon,\eta)$ by sampling from our prior $\pi(x)$ and computing the corresponding noise processes given that sample.

A given sample $x_s$ from our prior defines a specific noise configuration in our gate set.
Given that noise we can compute the expected output $p_s = \cA(x)$ using the exact forward model. 
This defines the effective bias, which are the Born probabilities, for multinomial outputs we will measure upon repeated sampling.
We sample from $\multinomial(p_s,N)$ a number of times to the equal experiment and store the statistics of $\epsilon$.
We use the same samples $x_s$ and $p_s$ to compute the corresponding sample of $\eta_s$. 
Computing the sample $\eta_s = p_s - (\bar m + \bar A x_s )$, we likewise add this to the samples of the full joint distribution. 
By stacking these samples of $x,\epsilon,\eta$ we can take empirically estimate the joint covariance and mean which we can then use in the \acronym{} protocol for this experimental setting. 
With access to full joint distribution we can determine the necessary components for the posterior computation, namely $\Gamma_{\eta}, \Gamma_{\epsilon}, \Gamma_{\eta x}, \Gamma_{\epsilon x}$ and $\bar\eta$.

\subsection{Posterior distribution}\label{subsection:methods-posterior distribution}

Now, by~\cref{eqn:bayesian-posterior}, we can determine the posterior $\pi(x|m)$ by applying the prior distribution. Since our prior $\pi(x)$ is multivariate Gaussian, it is conjugate~\cite{kaipio2007statistical} to our likelihood. 
This means that the posterior is likewise a multivariate Gaussian which we will denote $x|m \isnorm\left(\bar x_\mathrm{post},\Gamma_\mathrm{post}\right)$.
The mean $\bar x_\mathrm{post}$ is the solution to
\begin{align*}
\minimise{x}\|L_{e|x}(m-\bar{m}-\bar e|x - \bar A x)\|_2^2+\|L_{x} x\|_2^2\numberthis\label{eqn:posterior-mean-minimisation}
\end{align*}
and the covariance is
\begin{align}
    \Gamma_{x_\mathrm{post}} = \left(\Gamma_{x}^{-1} + \bar A\T \Gamma_{e|x}^{-1} \bar A\right)^{-1}.\label{eqn:posterior-covariance}
\end{align}
where $L_{e|x}$ and $L_{x}$ are the cholesky factors of $L_{e|x}\T L_{e|x} = \Gamma_{e|x}^{-1}$ and $L_{x}\T L_{x} = \Gamma_{x}^{-1}$, respectively.

Importantly, \cref{eqn:posterior-mean-minimisation} has a global minimum that can be efficiently computed using many approaches, including gradient-descent.
However, we can actually compute the minimiser for~\cref{eqn:posterior-mean-minimisation} in closed form.
It is simple to show that it the objective function~\cref{eqn:posterior-mean-minimisation} is identical to 
\begin{align}
    \minimise{x} \left\|\bar B x - y \right\|_2^2
\end{align}
where
\begin{align}
    \bar B :=& \begin{bmatrix}
    L_{e|x}(\bar A + \left( \Gamma_{\eta x} + \Gamma_{\epsilon x}\right)\Gamma_{x}^{-1})\\
    L_{x}
    \end{bmatrix}\\
    y:=&\begin{bmatrix}m-\bar m - \bar \eta|x\\
    0
    \end{bmatrix}.\label{eqn:psuedoinverse-operators}
\end{align}
This is now in the form of a linear least-squares objective where we can simply take
\begin{align}
    \bar x_\mathrm{post} = \left(\bar B\T\bar B\right)^{-1} \bar B\T y.
\end{align}

This defines the posterior which forms the basis of~\acronym{}.
However, the posterior distribution encodes our knowledge of the gate set given the data, we must also select a relevant point estimate from this distribution. The \emph{maximum a posteriori} estimator (the mode) is most commonly used, however, in general this will not be physical gate set, as desired. We will consider this in the next section.

\subsection{Physical priors and estimates} \label{section:physical-priors}

One of the shortcomings of the Gaussian prior model we introduced in~\cref{appendix:bayesian-model} is does not necessarily guarantee the underlying random variables will be physical.
The multivariate Gaussian priors are used to trade-off a physically-constrained prior for computational tractability. 
We can, however, use the convexity of the set of CPTP maps to our advantage to approximately update our prior to assign low probability to regions that are far from physical.

As motioned earlier, we use the PTM representation of channels.
For a CPTP map a PTM $\Lambda\in\mathbb{R}^{d^2\times d^2}$ takes the following form
\begin{align}\label{eqn:ptm-representation}
    \Lambda=\begin{bmatrix}
    1 & \vec{0}^T\\
    \vec{\tau}&\cU
    \end{bmatrix}.
\end{align}
The vector $\tau\in\mathbb{R}^{d^{2}-1}$ corresponds to the non-unital part of the channel and will be the zero vector for unital channels (map to the identity to the identity.)
The top row of the PTM is $\left[1,0,...,0\right]$ which corresponds to the trace-preservation condition of quantum processes.
This means that we can trivially impose the trace-preserving constraint in our prior by fixing these parameters, whilst also reducing the number of by parameters by $d^2$ per gate.
This means that these entries are completely determined in our prior and we only have to worry about complete-positivity, however this is not as visible a constraint for the PTM.
Complete positivity of a channel is equivalent to ensuring the Choi matrix is positive semi-definite. 

This means we need to solve a semi-definite program (SDP) to constrain our results to physical gates.
To project a sample gate from our prior, say $\Lambda_i$, onto the set CPTP gates we take
\begin{align*}
\hat\Lambda_i&=\argmin{\Lambda} \|\Lambda - \bar\Lambda_i\|_2\\
&\text{subject to\indent} \choi(\Lambda) \succeq 0.\numberthis\label{eqn:projected-estimator}
\end{align*} 
which can be efficiently solved using methods in convex optimisation~\citep{boyd2008graph}.
The SDPs in the results of this paper were computed using the CVX library~\citep{cvx}.
We use~\cref{eqn:projected-estimator} to project the MAP estimate to a point estimate for \acronym{}, which we will refer to as the \emph{projected maximum a posteriori} (PMAP) estimate. 
Similar approaches for projected estimators have been studied in the context of state tomography~\citep{gu2020faststate}.

for each channel $\Lambda_i$, which can be efficiently computed by a semi-definite program. In plain words we always take our estimator to be the closest physical channel to the mean, which will in general not be the mean. In the next section we consider how we can encode physical bounds on our prior distribution.

\section{Gate set results}\label{appendix:gate-set-results}

\begin{figure*}[b!]
    \centering
    \includegraphics[width=\linewidth]{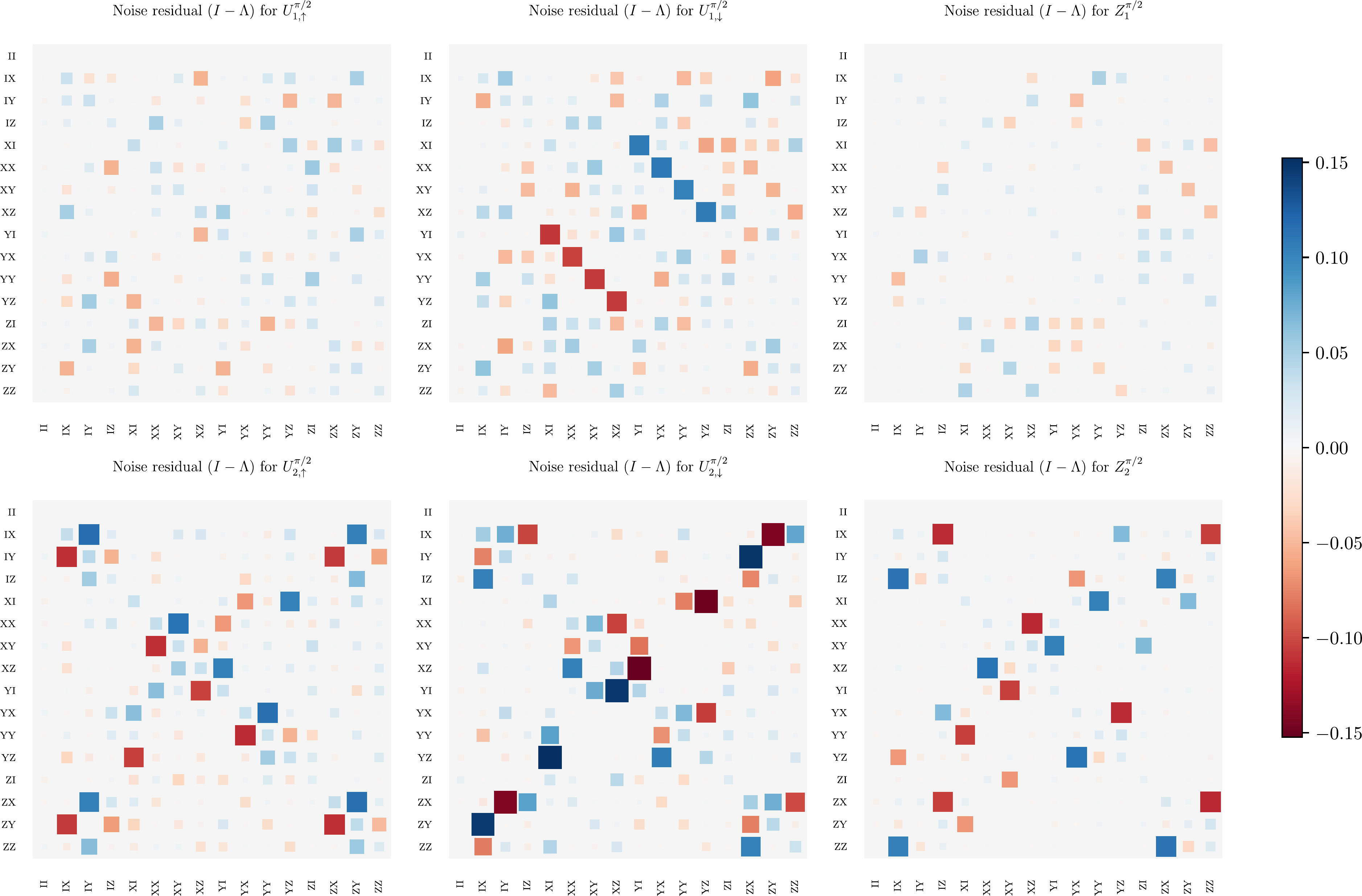}
    \caption{Gate set noise estimates presented as noise channel residuals $(I-\bar\Lambda)$.
    }
    \label{fig:gate-results-residuals}
\end{figure*}

\begin{figure*}
    \centering
    \includegraphics[width=\linewidth]{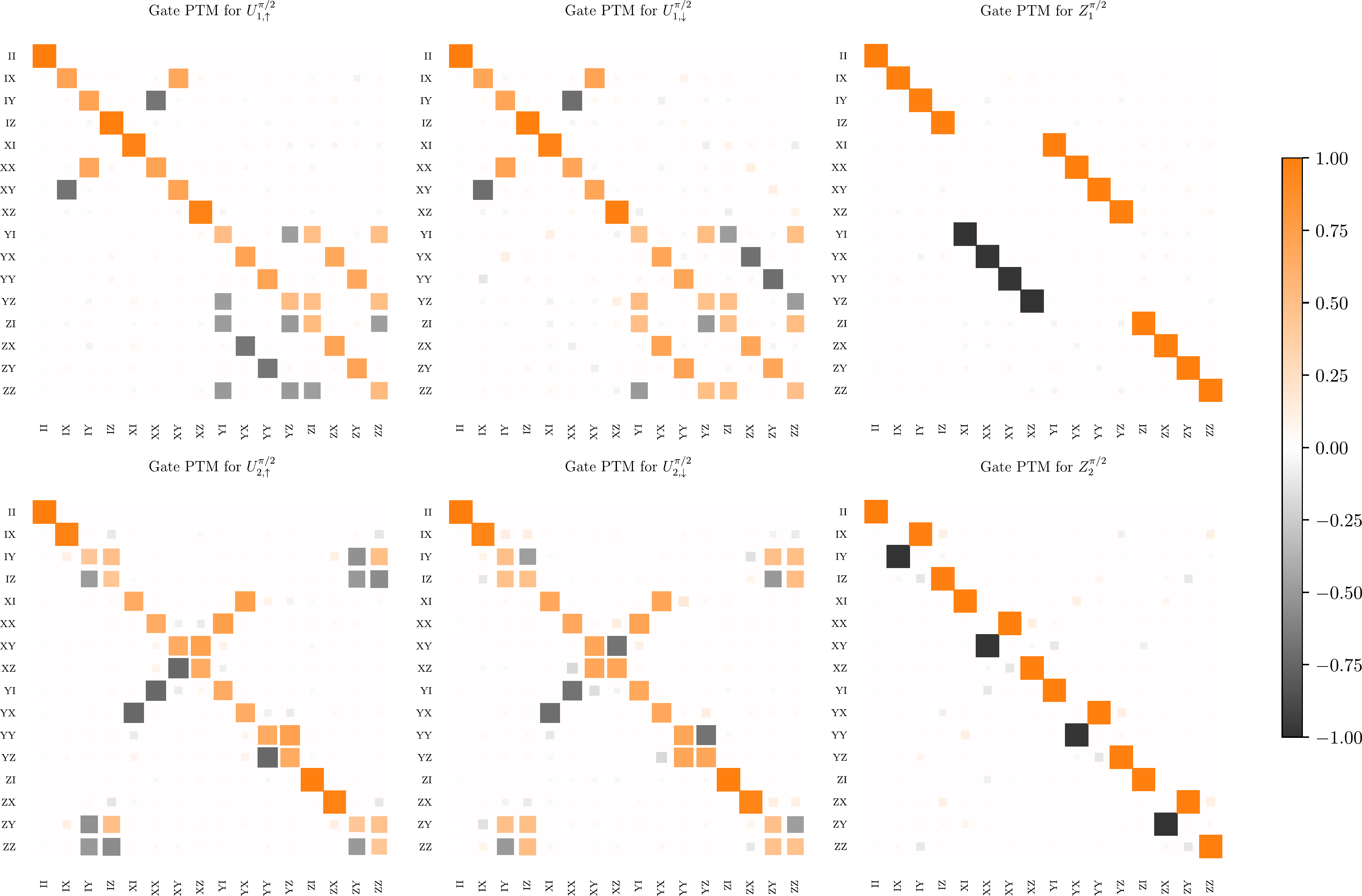}
    \caption{Gate set estimates presented as full gate Pauli transfer matrices. 
    }
    \label{fig:gate-results-ptms}
\end{figure*}

In this appendix we present the full results of all the channels from the gate set tomography in~\cref{section:results}. 
For each gate in the gate set we present the noise residual, $(I-\Lambda_G)$, in~\cref{fig:gate-results-residuals} and the full gate PTM, $\Lambda_G G$, in~\cref{fig:gate-results-ptms}.

\end{document}